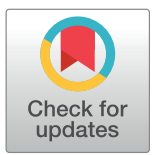

RESEARCH ARTICLE

# A model of octopus epidermis pattern mimicry mechanisms using inverse operation of the Turing reaction model

Takeshi Ishida*

Department of Ocean Mechanical Engineering, National Fisheries University, Shimonoseki, Yamaguchi, Japan

* ishida@fish-u.ac.jp

## Abstract

Many cephalopods such as octopi and squid can purposefully and rapidly change their skin color. Furthermore, it is widely known that some octopi have the ability to rapidly change the color and unevenness of their skin to mimic their surroundings. However, there has been little research published on the mechanisms by which an octopus recognizes its surrounding landscape and changes its skin pattern. We are unaware of any hypothetical model that explains this mimicry mechanism to date. In this study, the mechanism of octopus skin pattern change was assumed to be based on the Turing pattern model. Here, pattern formation using the Turing model was realized using an equivalent filter calculation model and a cellular automaton instead of directly solving the differential equations. It was shown that this model can create various patterns using two feature parameters. Furthermore, for visual recognition where two features are extracted from the Turing pattern image, a method that requires minimal calculation using the characteristics of the cellular Turing pattern model is proposed. These two calculations can be expressed in the same mathematical frame based on the cellular automaton model using a convolution filter. As a result, a model that is capable of extracting features from patterns and reconstructing those patterns rapidly can be created. This represents a basic model of the mimicry mechanism of octopi. Further, this study demonstrates the potential for creating a model with minimal learning calculation for application to machine learning.

## Introduction

The blue-ringed octopus (*Hapalochlaena fasciata*) is known for warning predators by producing a blue ring-shaped pattern on its body when it feels endangered. David Scheel et al. [1] reported that the octopus species *Octopus tetricus* uses body color patterns as signals in the context of hostile behavior within the species. These are examples of using epidermis patterns for communication purposes, such as intimidation. Furthermore, it is widely known that some octopi have the ability to purposefully and rapidly change the color and unevenness of their skin to mimic the surroundings [2, 3].

The mechanism used by these cephalopods to change skin color involves controlling the size of epidermal chromatophores connected to nerve cells. Chromatophores contain pigment granules that regulate color effects by changing in size. By adjusting the size of the chromatophore with muscles, it is possible to change the color rapidly. In addition, there are iridescent chromatophores, white chromatophores, and reflex cells below the layer of chromatophores that create a diverse color palate [4].

Although the mechanisms for the adjustment of skin color have been elucidated, there has been little research published on the mimicry mechanism by which the octopus recognizes the surrounding landscape and changes skin patterns accordingly. Presently, there is no published hypothetical model to explain this mimicry mechanism. There have been many papers regarding the high visual learning ability of octopi [5]. Considering these facts, it is appropriate to hypothesize that mimicry realized through chromatophores in the skin is controlled by visual processing of information in the brain and subsequent nervous system interaction with skin chromatophores. Regarding the texture of the skin, it can be inferred that information from the sensory organs of the tentacles is processed by the brain, leading to movement of the skin muscles. However, the mechanisms involved in this entire mimicry process have not yet been elucidated.

Furthermore, it is said that the eyes of an octopus can see only blue wavelengths [6]. However, because the retina is well developed, it is thought that light and shade patterns can be recognized. Considering a report that there are cells that sense light in the skin [7], it is possible that information from these cells and the eyes are combined to sense color. Much remains unclear regarding visual recognition of surroundings by octopi.

The purpose of this study is to construct a mathematical model of the mimicry process in octopi in which the pattern of the surrounding landscape is characterized visually and reproduced by patterns on the skin. This is a hypothetical model. However, just as the Turing model [8] contributed to the elucidation of the mechanisms of morphogenesis of living organisms, it is meaningful to construct a hypothetical model in order to elucidate the mechanisms of cephalopod mimicry.

It is generally said that the lifespan of an octopus is about one year [9]. It is therefore difficult for an octopus to learn mimicry through experience. The octopus is likely to have the ability to mimic instantly and autonomously.

A model of mimicry mechanisms can also be applied to machine learning, which responds to aggregate external information such as images. Processing with minimal learning is effective in expanding the range of use of artificial intelligence, and various studies regarding this are underway [10]. Engineering applications of artificial camouflage are presented in Li and Correll [11]. However, an effective mathematical model for cephalopod mimicry for engineering applications has not yet been reported.

Accordingly, an octopus mimicry model must be built from scratch. The following two models are given as the basis for constructing an octopus mimicry model:

1. A deep learning convolutional neural network (CNN) can be applied to the visual extraction of surrounding features. Because CNN [12] is based on visual recognition models, such as the Neocognitron [13], which applies the mechanism of the biological eye, it is appropriate for the models of the cephalopod eye.

2. Because the appearance of the skin pattern is expected to be consistent with the Turing pattern [8], the mechanism of pattern emergence according to the reaction-diffusion model (Turing model) can be applied to skin pattern creation. Long after the proposal of the Turing model, it has remained unknown whether a Turing pattern is really the same as the actual color pattern of an animal. The first observation of the formation of a Turing pattern in nature was conducted by Kondo and Asai who studied the stripes of the marine angelfish *Pomacanthus* [14]. Regarding hybrids, Miyazawa et al. [15] studied the color patterns of the purebred species and hybrids of salmonid fish and reported that each pattern could be explained by solving the Turing pattern model and reproduced from the intermediate value of the parameter of the purebred species. There are no reports of studies identifying octopus or squid patterns as Turing patterns, but striped or speckled patterns in animals appear to be typical Turing patterns.

This study is based on these two models, which appear to be completely different algorithms. However, as described in detail in Materials and methods, a cellular automaton model was constructed equivalent to the reaction-diffusion model modified as a filter calculation model that can reproduce patterns in the same mathematical frame as the convolutional operation of CNN.

In the extraction of feature parameters from patterns by CNN models, the back-propagation method is generally used to calculate the weight parameters. However, in this study, the possibility of feature measurement with a smaller amount of calculation by inverse calculation of the cellular automaton model was clarified. In this report, it is shown that both feature extraction from images and pattern reproduction can be handled using the same mathematical structure.

This model makes it possible to rapidly extract feature parameters from Turing patterns and reconstruct the Turing pattern from these feature parameters. Further biological studies are needed to determine if this model is consistent with the information processing mechanisms in cephalopods. This provides critical modeling in which visual information processing and the reaction process on the skin can be processed with a uniform integrated structure.

This is a novel hypothetical model for elucidating the mechanism of mimicry. Furthermore, minimization of calculation was demonstrated to determine the convolution filter for applications in machine learning, which will lead to the reduction of learning requirements for the CNN model.

## Materials and methods

### Outline of the model

In this study, the mechanism of octopus skin pattern formation was assumed to be based on the Turing model as shown in Fig 1. Here, pattern formation by the Turing model was realized by the equivalent filter calculation model using the cellular automaton instead of directly solving the differential equations. It was shown that this model can create various patterns with two feature parameters. Furthermore, a method was proposed wherein two features are extracted from the Turing pattern image that requires minimal calculation using the characteristics of the cellular Turing pattern model.

There are two possible models of camouflage. One model creates a camouflage pattern by the spatial arrangement of chromatophores; this model does not change the pattern for a long time. The second model creates a camouflage pattern by controlling the chromatophore within a short period. The octopus mimicry model proposed in this paper is the second model.

The first model can be formed by two types of morphogen models by the Turing model or a simple model by Reiter et al. [16]. In the Reiter et al. model, the relative positioning of two different types of chromatophores was examined in cuttlefish (*Sepia*), and a simplistic model was proposed based on the biological observations, whereby the each chromatophore of each color class defined a perimeter of inhibition around itself bounding the positioning of

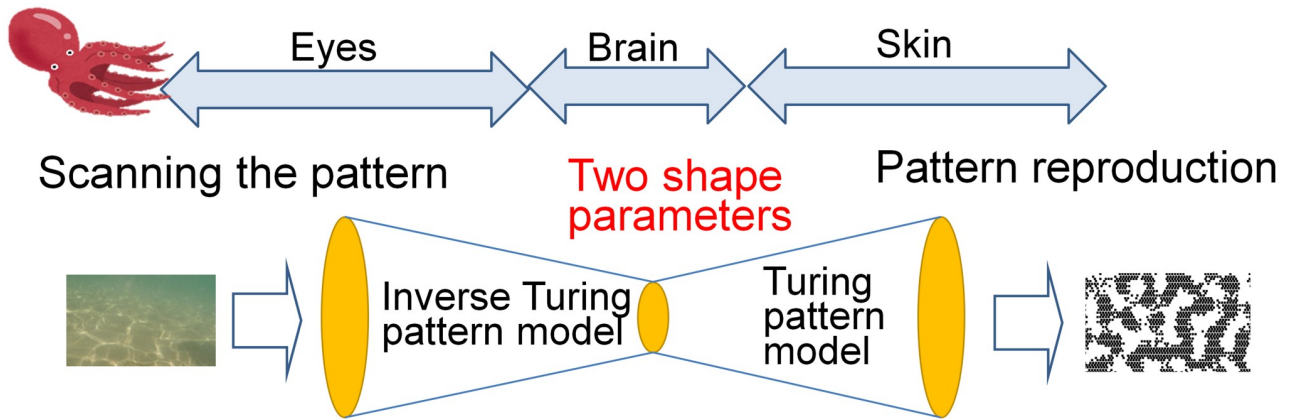

**Fig 1. Outline of the model of octopus epidermis pattern mimicry.** The mechanism of octopus skin pattern formation was assumed to be analogous to the Turing model and a cellular automaton. An inverse calculation model was proposed using the characteristics of the cellular Turing pattern model where two features are extracted from the Turing pattern image.

https://doi.org/10.1371/journal.pone.0256025.g001

new-born chromatophores. However, the second model cannot be explained by either of the models.

### Skin pattern model (Turing pattern model)

The Turing pattern model is a type of reaction-diffusion model that was introduced by Alan Turing in 1952 [8]. Therein, he treated morphogenesis as the interaction between activating and inhibiting factors. Typically, this model produces self-organized patterns through different diffusion coefficients of two morphogens, which are equivalent to activating and inhibiting factors. The general reaction-diffusion equations can be written as follows:

$$\begin{aligned}\frac{\partial u}{\partial t} &= d_1 \nabla^2 u + f(u, v), \\ \frac{\partial v}{\partial t} &= d_2 \nabla^2 v + g(u, v),\end{aligned} \quad (1)$$

where u and v are the morphogen concentrations, functions f and g are the reaction kinetics, and $d_1$ and $d_2$ are the diffusion coefficients. Previous studies have considered a range of functions f and g, and models such as the linear model, the Gierer–Meinhardt model [17], and the Gray–Scott model [18] have been used to produce typical Turing patterns.

Moreover, it is easy to solve such reaction-diffusion equations numerically and create Turing patterns in space using the difference method. However, deriving the parameters from the pattern by inversely calculating the differential equation generally requires many iterative operations and a large amount of calculation. Therefore, in this study, a method of estimating parameters based on a cellular automaton (CA) model that reproduces the Turing pattern instead of directly solving the differential equations was used.

CA models are discrete in both space and time. The state of the focal cell is determined by the states of the adjacent cells and the transition rules. The advantage of CA models is that they can describe systems that cannot be modeled using differential equations. Historically, various interesting CA patterns have been discovered. Markus et al. [19] demonstrated that a CA model could produce the same output as reaction-diffusion equations.

The Young's model [20] is a two-dimensional totalistic model that bridges reaction-diffusion equations and the CA model, which can produce Turing patterns. Adamatzky et al. [21]

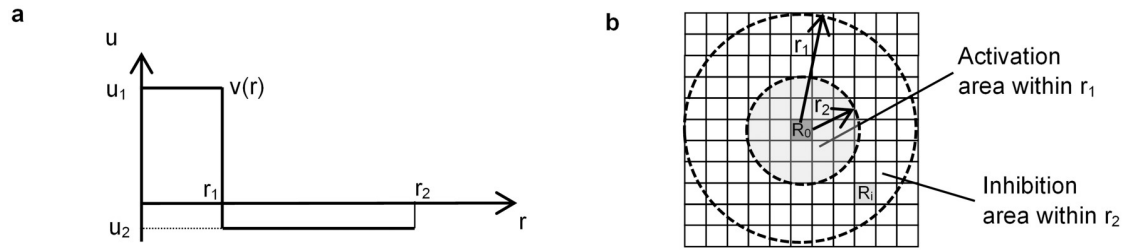

**Fig 2. Outline of Young's model.** (a) Function v(r) is a continuous step function representing the activation area and inhibition area. (b) The activation area has a radius $r_1$, and the inhibition area has an outer radius $r_2$.

studied a binary-cell-state eight-cell neighborhood two-dimensional CA model with semitotalistic transitions rules. Dormann et al. [22] also used a two-dimensional outer-totalistic model with three states to produce a Turing-like pattern. Tsai et al. [23] analyzed a self-replicating mechanism in Turing patterns with a minimal autocatalytic monomer–dimer system. These represent examples of models that produce Turing patterns.

The proposed model was based on a Young's CA model. Young's model uses a real number function (Fig 2a) to derive distance effects, with two constant values, $u_1$ (positive) and $u_2$ (negative). The calculation begins by randomly distributing black cells on a rectangular grid (Fig 2b). Then, for each cell at position $R_0$ in the two-dimensional field, the next cell state (black or white) of $R_0$ results from the sum of the function values of black cells at $R_i$ positions. $R_i$ is assumed to be a black cell within a radius $r_2$ from the $R_0$ cell, and i is the number of black cells within radius $r_2$ from $R_0$. The function v(r) is a continuous step function as shown in Fig 2a. The activation area, indicated by $u_1$, has a radius of $r_1$, and an inhibition area, indicated by $u_2$, has a radius of $r_2$ ($r_2 > r_1$) (Fig 2b). At position $R_0$, when $R_i$ is assumed to be a grid within $r_2$, function $v(|R_0 - R_i|)$ values are summed. Function $|R_0 - R_i|$ indicates the distance between $R_0$ and $R_i$. If $\sum_i v(|R_0 - R_i|) > 0$, the grid cell at point $R_0$ is marked as a black cell. If $\sum_i v(|R_0 - R_i|) < 0$, the grid cell becomes a white cell. If $\sum_i v(|R_0 - R_i|) = 0$, the grid cell does not change state [20]. Young showed that a Turing pattern can be generated using these functions. Spot patterns or striped patterns can be created with relative changes in $u_1$ and $u_2$.

In this Young's model, let $u_1 = 1$ and $u_2 = w$ (here $0 < w < 1$). If the state of the cell is set to 0 (white) or 1 (black), this model can be arranged as indicated below (2). The state of cell i is expressed as $c_i(t)$ ($c_i(t) = [0, 1]$) at time t. The following state $c_i(t + 1)$ at time t + 1 is determined by the states of the neighboring cells. Here, $N_1$ is the sum of the states of the domain within the $s_1$ mesh of the focal cell. Similarly, $N_2$ is the sum of the states of the domain within the $s_2$ mesh of the focal cell, assuming that $s_1 < s_2$.

$$N_1 = \sum_{i=1}^{S_1} c_i(t),$$
$$N_2 = \sum_{i=1}^{S_2} c_i(t). \quad (2)$$

Here, $S_1$ and $S_2$ are the numbers of cells within $s_1$ and $s_2$ meshes of focal cells, respectively. In addition, $s_2 = 2s_1$ was assumed in this paper. Fig 3 shows a schematic of the total sum of states $N_1$ and $N_2$. The next time-state of the focal cell is determined using expression (3). Here,

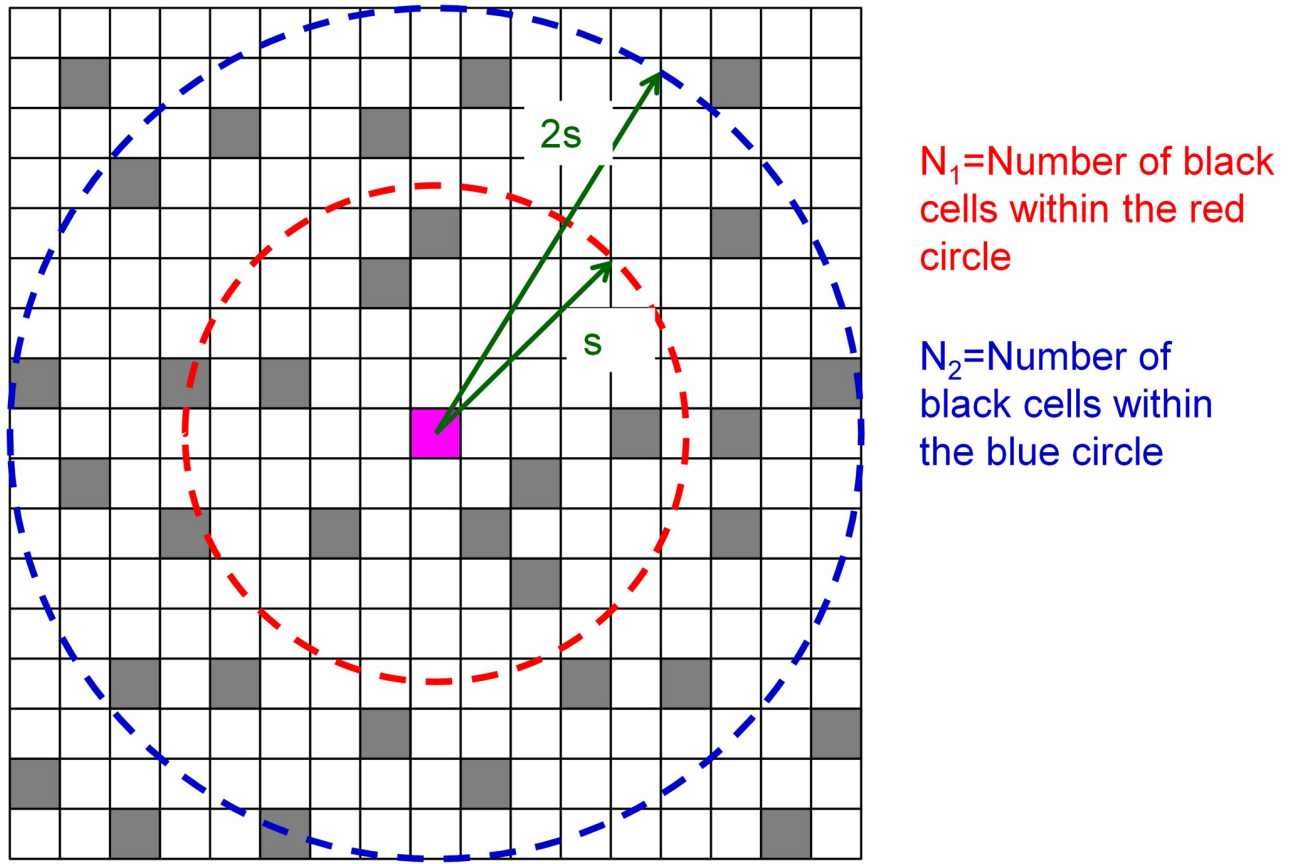

**Fig 3. Schematic diagram of the summation of states $N_1$ and $N_2$.** Each grid cell has state 0 (white) or 1 (black). The inner area has a domain within the s grid of the focal cell, and the outer area has a domain within the 2s grid.

https://doi.org/10.1371/journal.pone.0256025.g003

there are two parameters, namely, w and s, that determine the Turing pattern.

$$\text{Cell state at the next time step} = \begin{cases} 1 : \text{if } N_1 > N_2 \times w \\ \text{Unchanged} : \text{if } N_1 = N_2 \times w \\ 0 : \text{if } N_1 < N_2 \times w \end{cases}. \qquad (3)$$

Furthermore, the operation of counting state-1 cells as $N_1$ and $N_2$ is equivalent to the convolution operation, which uses a filter with an s range of 1 and a 2s range of w (shown in Fig 4) to sum the values obtained by the AND calculation of the filter and mesh space. This process is equivalent to CNN in performing convolution operations on the mesh space.

The essence of the Turing pattern model is that the short- and long-range spatial scales are each affected by separate factors, and pattern formation emerges from nonlinear interactions between the two factors. Turing used two chemicals with different diffusion coefficients to create these short- and long-range spatial effects. However, as long as a difference exists between long- and short-range effects, other implementation methods can be applied. This model used two ranges of s mesh and 2s mesh to create a difference effect. This model is therefore essentially the same as a reaction-diffusion model.

On the other hand, in the skin pattern model (Turing pattern model) of this paper, the color of a specific cell, white or black, is determined by the proportion of black cells in the

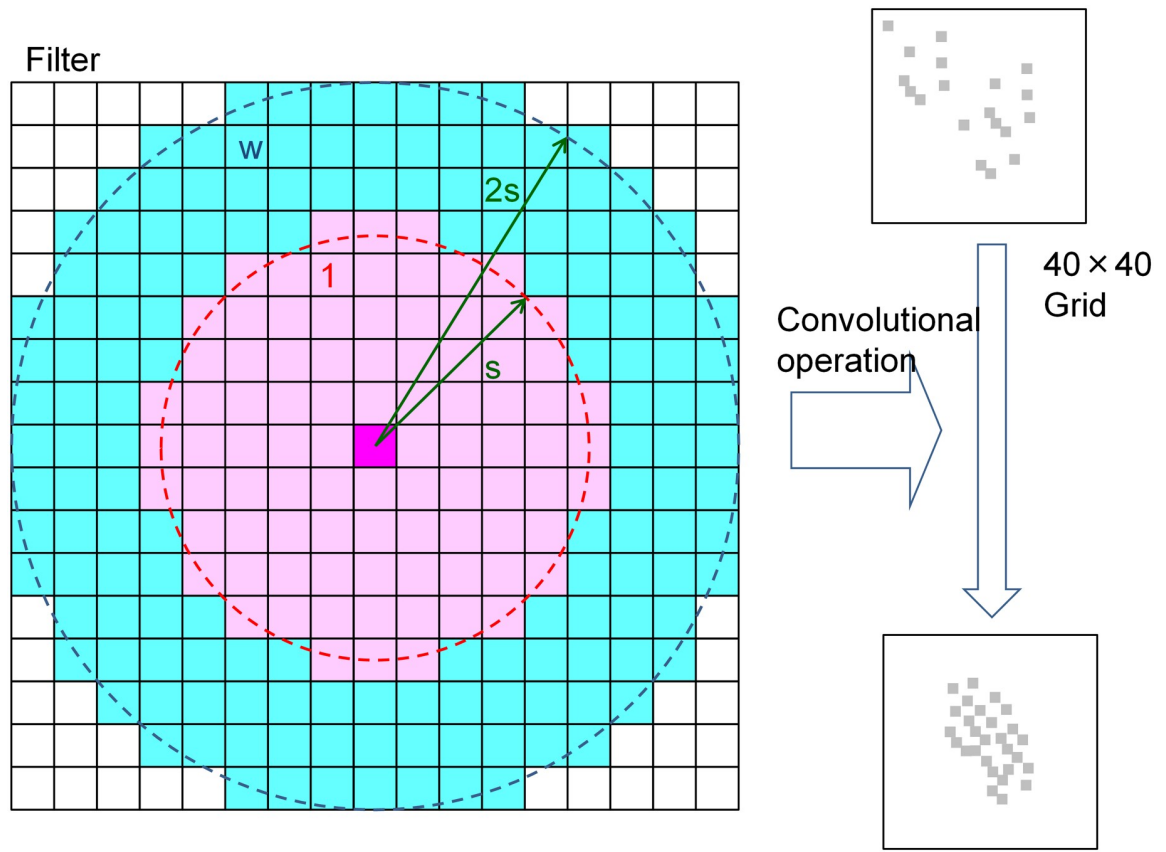

**Fig 4. Overview of convolution operation by the filter.**

https://doi.org/10.1371/journal.pone.0256025.g004

surrounding double range using the conversion method by Young. It is only necessary to observe the state of the neighborhood of a specific cell. If the unit composed of the chromatophore and the muscle cell is connected to the neuron-like nerve cell, they can change the state of their chromatophores in a short period. This way, it is possible to describe the patterns that change in such a short time with my model.

### Information processing model of visual recognition in the octopus: A model to extract shape parameters by inverse Turing operation

A general method for extracting feature parameters from patterns is to build a predictive model from training data consisting of images and feature parameters by machine learning, such as with a CNN model. However, this method requires a large amount of training data and calculation because of the back-propagation method. It is unlikely that octopi use the same method as it naturally possesses the ability to mimic without learning. There may be a mechanism for acquiring feature parameters rapidly without learning or numerous optimization calculations.

In this study, the feature parameters were calculated by back-calculation using the CA model to create the Turing pattern described in the previous section. However, parameter values cannot be calculated by simply performing the reverse operation. This is because the reverse process of counting state 1 in a certain range cannot be determined in a single way. In this study, the method described below was devised.

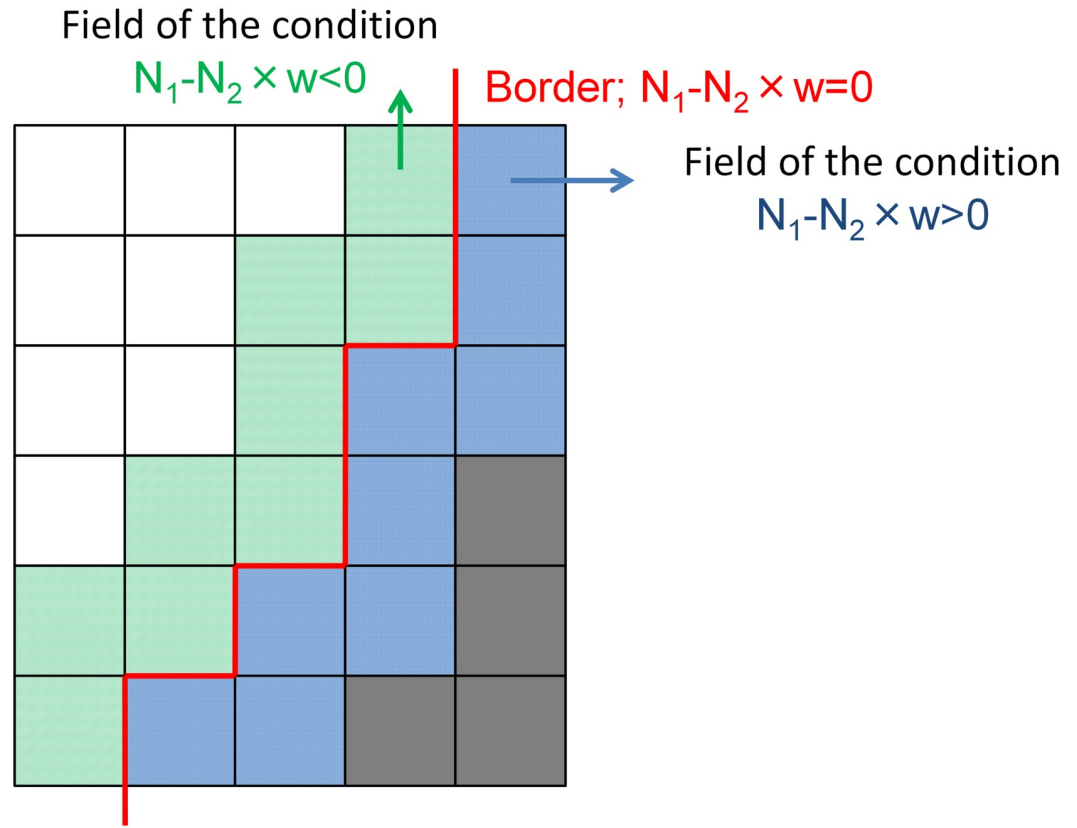

**Fig 5. An example of the boundary between black cells (state 1) and white cells (state 0).** In the convolution calculation with the filter on the black cells (1) in contact with the white cells (0), the value of $N_1 - N_2 w$ is $>0$. In the same calculation on the white cells (0) in contact with black cells (1), the value of $N_1 - N_2 w$ is $<0$.

https://doi.org/10.1371/journal.pone.0256025.g005

In the CA model algorithm that can produce the Turing pattern described in the previous section, $N_1 = N_2 \times w$ holds in the edge lines of the patterns. Therefore, it is possible to calculate w from the ratio of $N_1$ and $N_2$ in the cells on the pattern edge lines.

$$w = N_1/N_2 \tag{4}$$

However, the values of $N_1$ and $N_2$ change with value s, and value w on the pattern edges also changes with s. When s is properly determined, $N_1/N_2$ (= w) matches on all pattern edges, and w is also specified, but when s is different from the true value, the values of w are not fixed on the pattern edges. Conversely, if s is adjusted to the true value, the deviation of the values of w on the edges disappears and w becomes constant so that it is possible to determine s and w at the same time.

As shown in Fig 5, as the space is discrete on the edges of patterns where the white and black cells are in contact, there are no cells in which the value w is equal to $N_1/N_2$. The w values are $<N_1/N_2$ in the black cells at the pattern edges, and the w values are $<N_1/N_2$ in the white cells at the pattern edges. Therefore, w was determined in this study by averaging the w values of the black-and-white cells at all edges of patterns.

Further, "determining s so that all w values on the edges are equal" means that the deviation of the w values on the edges is minimized. In this model, an equivalent evaluation was used: "The ratio of cells with $w < N_1/N_2$ is at a maximum on black boundary cells, and the ratio of cells with $w > N_1/N_2$ is at a maximum on white boundary cells." That is, when the following index is maximized, s and w are the values to be obtained.

$$\text{Index} = \frac{\alpha}{\alpha_t} + \frac{\beta}{\beta_t}, \tag{5}$$

α: Number of black cells on the edge lines where $w < N_1/N_2$
$\alpha_t$: Number of black cells on the edge lines
β: Number of white cells on the edge lines where $w > N_1/N_2$
$\beta_t$: Number of white cells on the boundary.

As a specific calculation method, s and w were specified by calculation of the index value, which starts at a large initial value s, and by finding the s value at which the index becomes maximum while gradually decreasing s. This is a much smaller amount of computation than is required for the back-propagation method using a neural network. Fig 6 describes this calculation model.

## Model implementation and calculation conditions

The calculation program was implemented in Python. The program is divided into three parts: the first part that sets the w and s parameters and calculates the Turing pattern, the second part that estimates the w and s parameters from the obtained Turing pattern image, and the third part that calculates the Turing pattern again from the estimated w and s values. The grid was a two-dimensional area of 40 × 40 cells, and periodic boundary conditions were used to calculate the boundary area. The calculations were carried out where w = 0.175, 0.20, 0.25, 0.30, and 0.325 and s = 3, 4, 5, 6, 7, and 8. In this paper, only Turing patterns were focused.

Furthermore, actual images were applied to the inverse Turing calculation to evaluate the proposed model's applicability to non-Turing pattern backgrounds, such as the seabed encountered by octopuses. A photograph of a rocky coastal area resembling the background of the seabed encountered by octopus was taken and processed into a black-and-white binary image. Then, the parameters were extracted using the algorithm from the present proposal, and the image was reproduced using those parameters.

## Results

### Results of the Turing pattern-producing model

The emergent pattern was calculated using w and s parameters and the skin pattern formation model based on the Young's model. Fig 7 shows the results of the model using expression (1) and a rectangular grid. For initial conditions, state-1 cells were set randomly in the 40 × 40 calculation field. Turing-like patterns emerged as w and s parameters changed. When w was relatively large, spot patterns were observed, whereas in the middle range of w values, stripe patterns emerged. When w was 0.15 or lower, all cells in the field had a value of 1 (shown in black). Changing the s parameter merely changed the scale of the patterns.

The calculation starts in a random state and settles with a steady pattern in about 10 timesteps. Because 0 and 1 are set randomly in the initial arrangement, the pattern was not exactly the same for each run, even with the same parameters. However, the pattern characteristics (the width of the striped pattern and size of the speckled pattern) were dependent only on the parameters.

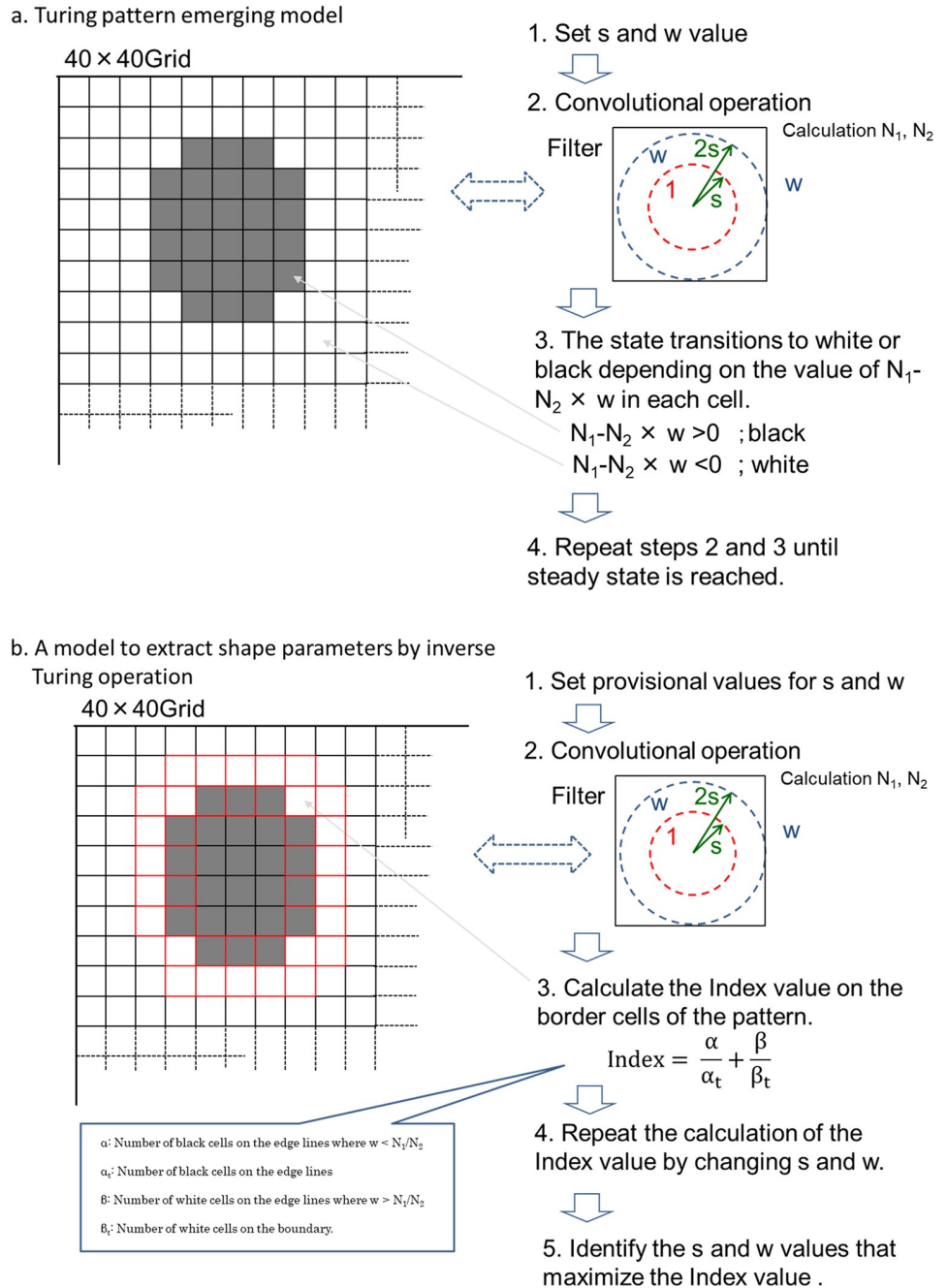

**Fig 6. Diagram of pattern emergence calculation and its inverse calculation process.** (a) The specific pattern emergence model by filter convolution calculation from two parameters, s and w. (b) The parameters s and w by the inverse calculation model.

https://doi.org/10.1371/journal.pone.0256025.g006

### Identification of feature parameters by inverse Turing calculation

Figs 8–11 show calculation results using feature parameters s and w and inverse Turing operation. The pattern reproduction results were based on the estimated parameters. The image calculated from the pattern-producing model, which was the result of estimating the s and w

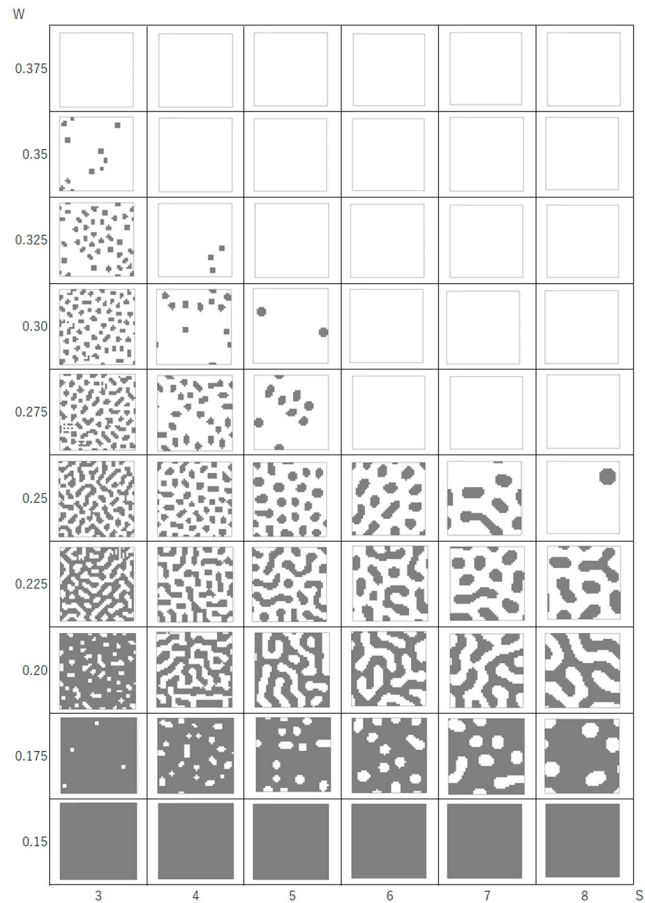

**Fig 7. Results for the model using expression (1) in a square grid.** S values are represented on the horizontal axis, and w values are represented on the vertical axis.

https://doi.org/10.1371/journal.pone.0256025.g007

parameters from the image, and the reconstructed image are shown under each condition in the figures.

From these results, it can be seen that image reproducibility was good under most calculation conditions. However, when w and s values were low, forming a small white spot pattern on a black background, the reproducibility of the image was poor. This may have been due to the large error in the value of the "index" when there were only a few small white spots on a black background. The same occurred when there were a small number of small black spots on a white background.

In the pattern-producing model, the pattern changed rapidly from a striped pattern to a solid black color when the w value was 0.25 or less. As shown in Fig 7, the pattern changed from speckled to a black background when w was between 0.175 and 0.15. This caused errors in the index, and a slight difference in the estimated w value caused a large change in the reproduced pattern image.

Table 1 shows the results of calculating the index values from the images created by each w value at s = 5.0. Index values were calculated or each w value pair with s values of up to 8. The index value was at a maximum at an s value of 5.0, except when the w value was 0.175. The value of s could be specified by the calculation method from the index value proposed in this study. In addition, the calculation for finding the parameters was determined by iterative

| Set value | | Image calculated with the set value | Predicted value | | Image calculated with predicted value | Reproducibility of image |
| --- | --- | --- | --- | --- | --- | --- |
| s | w | | s | w | | |
| 8 | 0.25 | 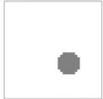 | 7.5 | 0.243 | 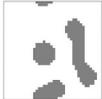 | ○ |
| 8 | 0.225 | 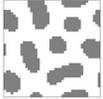 | 7.5 | 0.224 | 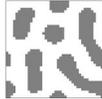 | ○ |
| 8 | 0.20 | 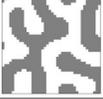 | 7.5 | 0.201 | 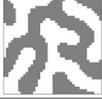 | ○ |
| 8 | 0.175 | 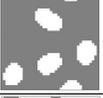 | 7.5 | 0.179 | 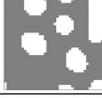 | ○ |
| 7 | 0.25 | 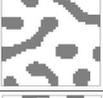 | 7.0 | 0.244 | 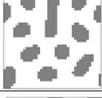 | ○ |
| 7 | 0.225 | 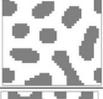 | 7.0 | 0.218 | 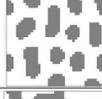 | ○ |
| 7 | 0.20 | 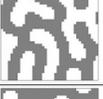 | 7.0 | 0.200 | 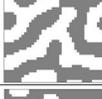 | ○ |
| 7 | 0.175 | 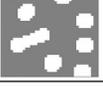 | 7.0 | 0.175 | 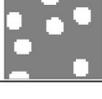 | ○ |

**Fig 8. Calculation results (1).**

https://doi.org/10.1371/journal.pone.0256025.g008

calculation using several steps, which took several seconds even on a personal computer not using a standard GPU.

At the rocky shore in front of the National Fisheries University in the city of Shimonoseki in Japan (Fig 12a), the rocks below the sea surface at high tide were photographed (Fig 12b) at the shooting range of approximately 1 square m. A part of this photograph was converted into a black-and-white binary image (40 × 40 mesh) by brightness (256 layers). The thresholds for separating white and black were set to 100 and 25, respectively, and two types of images were obtained (Fig 12c and 12d). Then, the algorithm of the present proposal was applied to extract the parameters. The images were then reproduced using the extracted parameters (Fig 13).

## Discussion

It was possible to infer feature parameters from Turing patterns and reconstruct similar Turing patterns under specific and controllable conditions using the proposed model. Unlike in the CNN model, it was not necessary to perform a large amount of calculation to determine feature parameters.

This model will serve as a hypothetical model for mechanisms of cephalopod mimicry and will be useful for future biological research. However, it has been pointed out that, among the various patterns in nature, the patterns according to the Turing model are only observed in a

| Set value | | Image calculated with the set value | Predicted value | | Image calculated with predicted value | Reproducibility of image |
|---|---|---|---|---|---|---|
| s | w | | s | w | | |
| 6 | 0.25 | | 6.0 | 0.239 | | ○ |
| 6 | 0.225 | | 6.0 | 0.218 | | ○ |
| 6 | 0.20 | | 6.0 | 0.199 | | ○ |
| 6 | 0.175 | | 6.0 | 0.178 | | ○ |
| 5 | 0.275 | | 5.0 | 0.268 | | ○ |
| 5 | 0.25 | | 5.0 | 0.236 | | ○ |
| 5 | 0.225 | | 5.0 | 0.216 | | ○ |
| 5 | 0.20 | | 5.0 | 0.199 | | ○ |
| 5 | 0.175 | | 4.5 | 0.206 | | × |

**Fig 9. Calculation results (2).**

https://doi.org/10.1371/journal.pone.0256025.g009

limited part of the world. Patterns on the seabed are not necessarily analogous to Turing patterns. In this paper, although only one image was used, it appeared possible to reproduce the black-and-white ratio by processing the photographs of the rocks. In the future, it is necessary to examine whether various non-Turing patterns can be reproduced with this model. To describe the possibilities of this point, the model proposed in the present study can generate patterns that cannot be created by numerically solving Turing's partial differential equations. For example, for the following part in equation (3):

$$\text{Unchanged}: \text{if} \quad N_1 = N_2 \times w,$$

by expanding the unchanged area of this part as shown below, patterns such as the one shown in Fig 14 can be created.

$$\text{Unchanged}: \text{if} \ (N_1 \geq N_2 \times w - 2.0) \ and \ ((N_1 \leq N_2 \times w + 2.0)$$

Here, 2.0 in the equation is an arbitrary value. As this number increases, the boundary of the Turing pattern becomes blurred. Thus, more diverse patterns can be considered for this model in the future.

| Set value | | Image calculated with the set value | Predicted value | | Image calculated with predicted value | Reproducibility of image |
| --- | --- | --- | --- | --- | --- | --- |
| s | w | | s | w | | |
| 4 | 0.30 | | 4.0 | 0.266 | | × |
| 4 | 0.275 | | 4.0 | 0.243 | | △ |
| 4 | 0.25 | | 4.0 | 0.227 | | ○ |
| 4 | 0.225 | | 4.0 | 0.215 | | ○ |
| 4 | 0.20 | | 4.0 | 0.198 | | ○ |
| 4 | 0.175 | | 3.0 | 0.213 | | × |

**Fig 10. Calculation results (3).**

https://doi.org/10.1371/journal.pone.0256025.g010

| Set value | | Image calculated with the set value | Predicted value | | Image calculated with predicted value | Reproducibility of image |
| --- | --- | --- | --- | --- | --- | --- |
| s | w | | s | w | | |
| 3 | 0.325 | | 3.5 | 0.231 | | × |
| 3 | 0.30 | | 3.0 | 0.253 | | △ |
| 3 | 0.275 | | 3.5 | 0.204 | | ○ |
| 3 | 0.25 | | 3.5 | 0.195 | | ○ |
| 3 | 0.225 | | 3.0 | 0.225 | | ○ |
| 3 | 0.20 | | 3.0 | 0.216 | | ○ |
| 3 | 0.175 | | 3.5 | 0.169 | | × |

**Fig 11. Calculation results (4).**

https://doi.org/10.1371/journal.pone.0256025.g011

Table 1. The results of calculating the index values from images created by a range of paired w and s values, peaking at s = 4.5–5.

|   |   | w |   |   |   |   |
|---|---|---|---|---|---|---|
|   |   | 0.275 | 0.25 | 0.225 | 0.2 | 0.175 |
| s | 7.5 | 1.615 | 1.502 | 1.441 | 1.536 | 1.637 |
|   | 7.0 | 1.704 | 1.723 | 1.660 | 1.703 | 1.800 |
|   | 6.5 | 1.693 | 1.758 | 1.689 | 1.737 | 1.795 |
|   | 6.0 | 1.845 | 1.857 | 1.847 | 1.872 | 1.869 |
|   | 5.5 | 1.960 | 1.941 | 1.940 | 1.968 | 1.929 |
|   | 5.0 | **1.971** | **1.950** | **1.968** | **1.990** | 1.940 |
|   | 4.5 | 1.966 | 1.921 | 1.951 | 1.986 | **1.995** |
|   | 4.0 | 1.793 | 1.867 | 1.895 | 1.894 | 1.931 |
|   | 3.5 | 1.845 | 1.874 | 1.871 | 1.903 | 1.973 |
|   | 3.0 | 1.833 | 1.838 | 1.795 | 1.861 | 1.973 |

https://doi.org/10.1371/journal.pone.0256025.t001

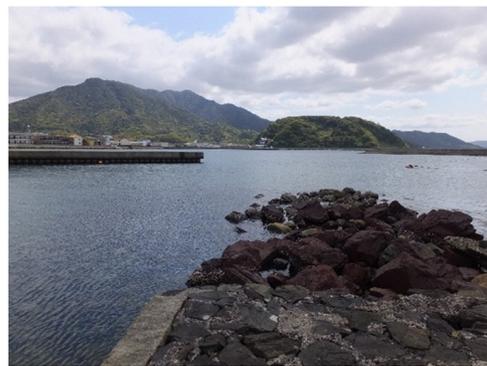

(a) Photographed beach(Shimonoseki City of Japan, the coast in front of National Fisheries University)

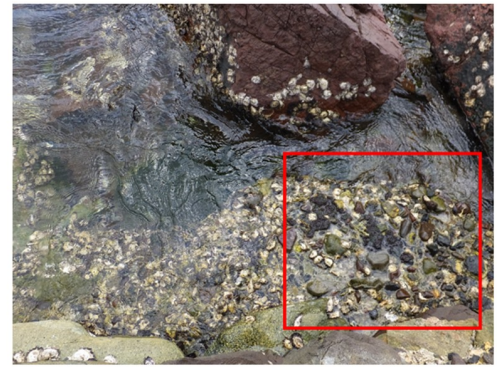

(b) Rocky photo for image-processing

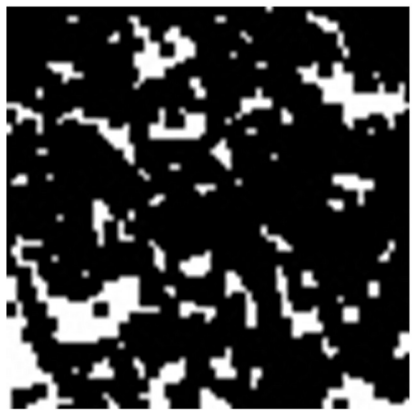

(c) Binarized image (threshold value 100): Black ratio = 82.7%

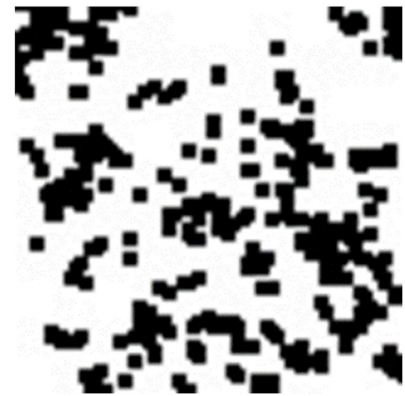

(d) Binarized image (threshold value 25): Black ratio =19.1%

**Fig 12. An image of coastal rocks for image-processing.**

https://doi.org/10.1371/journal.pone.0256025.g012

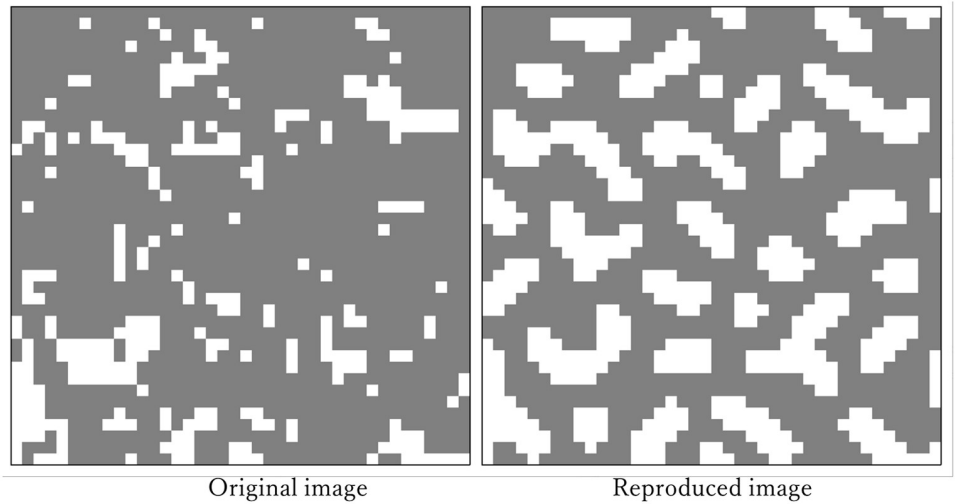

(a) Reproduced image from rocky binarized image (threshold value 100)

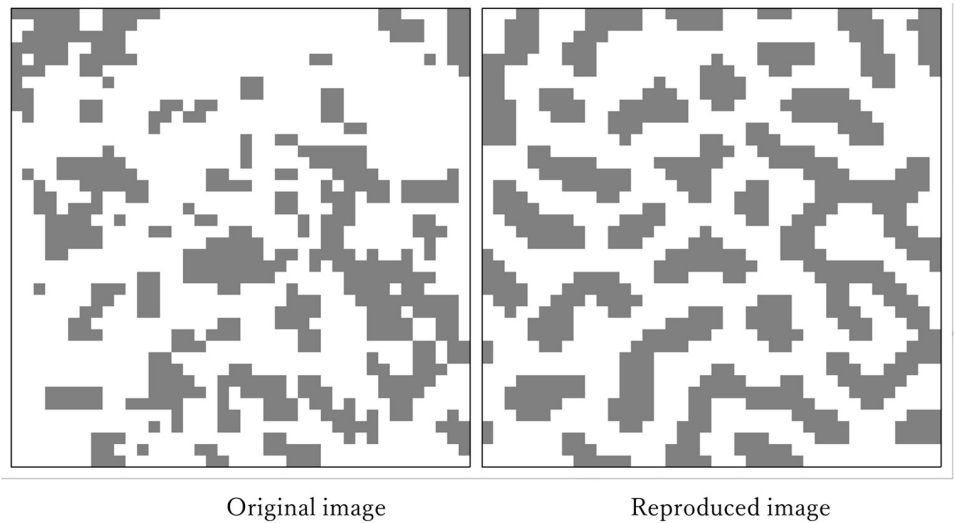

(b) Reproduced image from rocky binarized image (threshold value 25)

**Fig 13. Results of the reproduction of the coastal rocks from binary images using the proposed algorithm.**

https://doi.org/10.1371/journal.pone.0256025.g013

In addition, the filter shape was a perfect circle in this model (Fig 3). If a distorted elliptical filter is used, it is thought that more diverse shapes can be created. Therefore, it is necessary to examine various filter shapes.

This model clarified that it is possible to reproduce a Turing pattern processed with a perfect circle filter and that it is possible to create a model that can recognize a Turing pattern with minimal learning. At present, it is difficult to distinguish various geometric shapes, such as circles and squares, in this model. Although this model is inferior for reconstructing a detailed shape, it is a suitable model for rapid processing and reconstruction of specific patterns in the surrounding space.

In this regard, it is suggested that the fine examination of cephalopod skin patterns does not transition smoothly in the (s, w) plane (Fig 7). The calculation of the present model does not

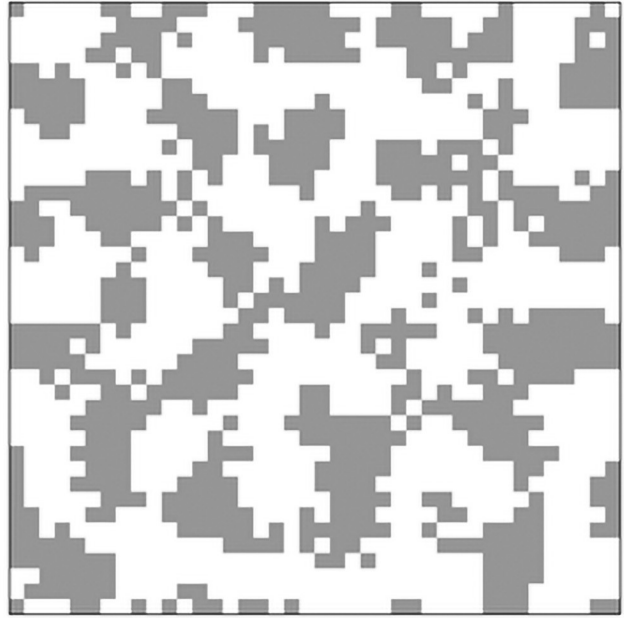

**Fig 14. Example of pattern emergence using a modified skin pattern model (calculation results under the conditions of s = 5.0, w = 0.225).**

https://doi.org/10.1371/journal.pone.0256025.g014

involve a broad continuous search of the s and w planes. Therefore, the pattern of the octopus skin does not continuously change in the space shown in Fig 7. This model proposes a method to extract a specific s and w within a short time for patterns seen by the octopus.

By clarifying which patterns can be reproduced using the proposed model and which pattern cannot, it is possible to develop an experimental method to test this hypothesis. For example, the checkerboard pattern shown below is considered a pattern that cannot be reproduced by the Turing model. It is believed that the model can be evaluated by clarifying patterns that cannot be reproduced with this model and testing whether these patterns can be reproduced with cephalopods. It can be seen from the reproduction of the patterns from some checkered patterns using this model (Fig 15) that none of the patterns can be reproduced properly. The experiment that observes the mimicry pattern when an octopus is placed on the background of a checkered pattern may be one of the methods to verify this model. Chiao and Hanlon reported experiments in cuttlefish using a checkerboard pattern in the background [24], and it is thought that they are trying to approximate a Turing-based speckled pattern. To clarify the mechanism underlying the expression of these patterns, it is necessary to conduct a variety of experiments.

In addition, the body pattern of the peacock flounder (Bothus mancus) appears to be a checkerboard pattern. However, upon closer inspection, the two types of black and white spots appear to be shifted and superimposed. It is thought that the same pattern can be constructed in this model by constructing a two-color model, thus further investigation is necessary.

In addition, there is a suggestion that patterns are made up of loosely defined subcomponents (e.g., linear borders and spots at predetermined positions), which form patchworks of shapes at certain scales. To address this point, the computational model assumed that each cell is homogeneously connected to the surrounding cells. A simple model of the composition of the cephalopod epidermis is shown in Fig 16. Neurons can be modeled as being connected to neurons of nearby cells. The parameter, s, in this model corresponds to the parameter of the

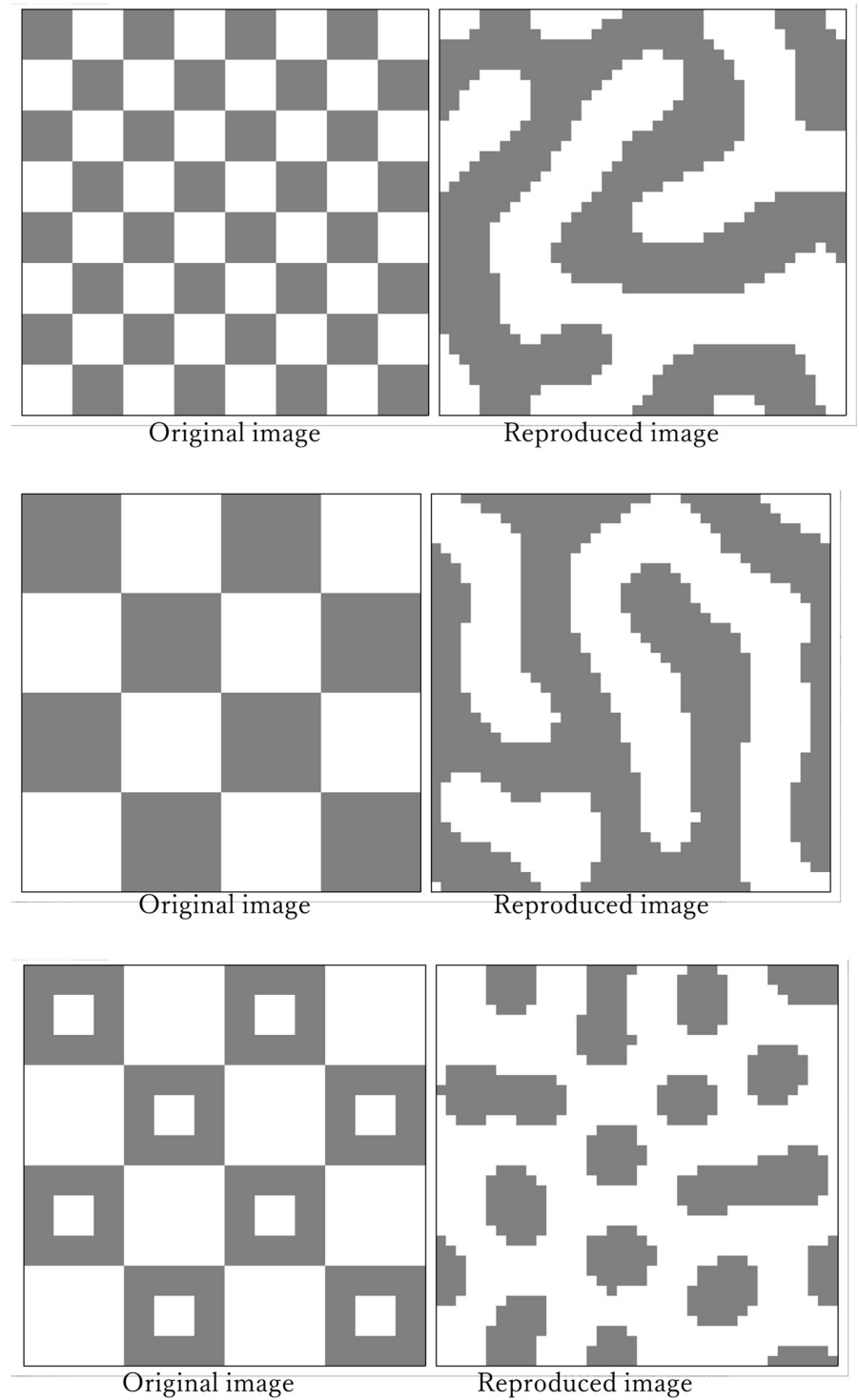

**Fig 15. Reproduced results of the checkerboard pattern.**

https://doi.org/10.1371/journal.pone.0256025.g015

range of information transmitted from distant cells. The parameter, w, is considered equivalent to the weight of moving muscle cells relative to the total information transmitted from the neighbor. When the muscle cells move, the pigment cells are stimulated and the pattern changes. On the other hand, the actual neural model of the cephalopod epidermis is not

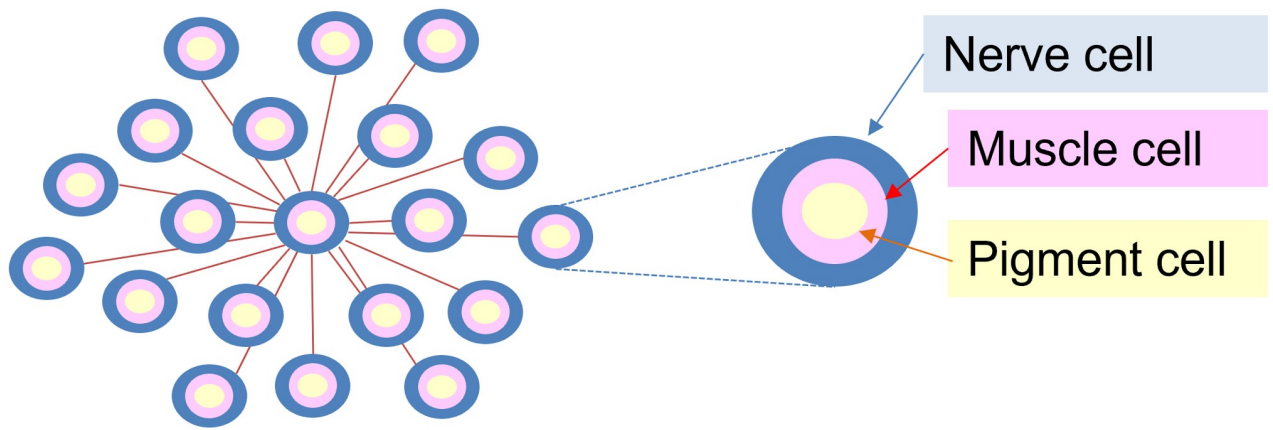

**Fig 16. Epidermal cell composition model.**

https://doi.org/10.1371/journal.pone.0256025.g016

necessarily homogeneous, and there may be substructures within the neural connections. This point needs to be further examined.

As shown in Fig 17, for applications to machine learning, it is possible that the discrimination performance will be improved by preparing directional filters, such as elliptical filters.

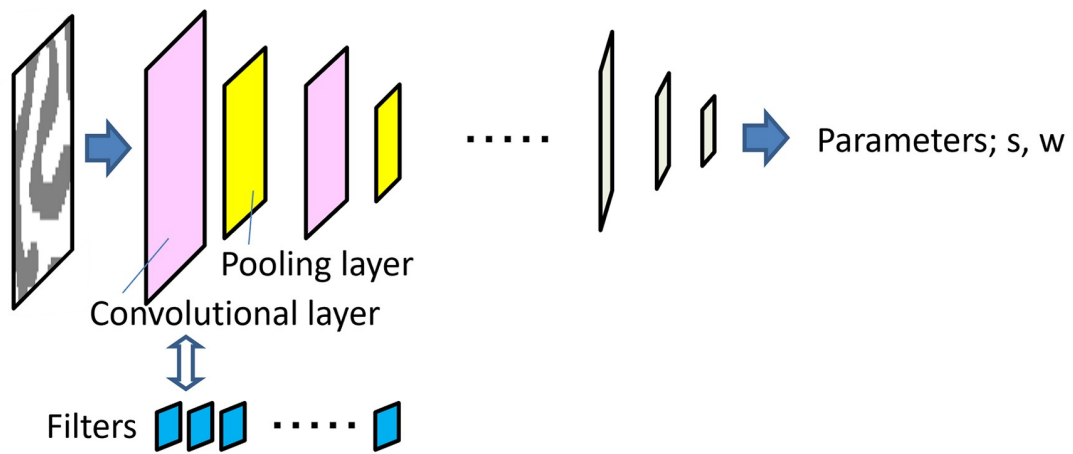

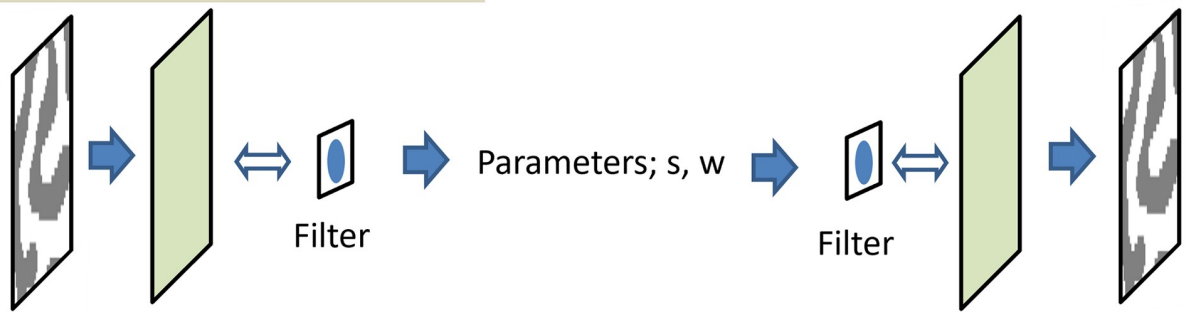

**Fig 17. Comparison between the general convolutional neural network and the proposed model.**

https://doi.org/10.1371/journal.pone.0256025.g017

Even in the CNN model, it is common to use dozens of filters, and it is thought that cognitive ability is enhanced by combining various anisotropic filter patterns. Furthermore, by integrating a CNN model, which learns the weights of the filter from random values, and this model, which uses a specific pattern filter, it may be possible to construct a new image recognition model with minimal learning and high recognition ability. With the CNN model, even simple image discriminations like those described in this paper would take several minutes on a typical desktop computer. However, with the present model, the calculation can be completed in the order of several seconds.

## Conclusions

In this study, a reaction-diffusion CA model that generates Turing patterns was constructed, and a mathematical model that extracts feature parameters from Turing pattern images was constructed. We demonstrated the combination of these two calculations expressed in the same mathematical frame based on a CA model using a convolution filter.

As a result, a model capable of extracting features from patterns and reconstructing them rapidly was created. This is a basic model for the mimicry mechanisms of the octopus. In addition, the possibility of creating a model for machine learning with minimal learning calculation was shown.

## Supporting information

**S1 File.**
(PY)

**S2 File.**
(XLSX)

## Acknowledgments

The authors would like to thank Enago ([www.enago.jp](www.enago.jp)) for the English language review.

## Author Contributions

**Conceptualization:** Takeshi Ishida.

**Data curation:** Takeshi Ishida.

**Formal analysis:** Takeshi Ishida.

**Funding acquisition:** Takeshi Ishida.

**Investigation:** Takeshi Ishida.

**Methodology:** Takeshi Ishida.

**Project administration:** Takeshi Ishida.

**Resources:** Takeshi Ishida.

**Software:** Takeshi Ishida.

**Supervision:** Takeshi Ishida.

**Validation:** Takeshi Ishida.

**Visualization:** Takeshi Ishida.

**Writing – original draft:** Takeshi Ishida.

**Writing – review & editing:** Takeshi Ishida.